\documentclass{amsart}

\usepackage{amsfonts, amsmath, amssymb, amsthm, bbm}
\usepackage{hyperref}

\usepackage[
backend=biber,
style=alphabetic,
citestyle=alphabetic
]{biblatex}
\usepackage{calc}
\usepackage[T1]{fontenc}
\usepackage[a4paper, total={6in, 8in}]{geometry}
\usepackage{graphicx}
\usepackage{hhline}
\usepackage[utf8]{inputenc}
\usepackage{multirow}
\usepackage{setspace}
\usepackage{verbatim}
\usepackage{float}

\theoremstyle{plain}
\newtheorem{theorem}{Theorem}[section]

\newtheorem{remark}[theorem]{Remark}

\newcommand{\bft}[1]{\boldsymbol{f}_t}

\newcommand{\bsigmat}[1]{\boldsymbol{\sigma}_t}
\renewcommand{\d}{\,\mathrm{d}}

\newcommand{\E}{\mathbb{E}}

\renewcommand{\P}{\mathbb{P}}

\newcommand{\R}{\mathbb{R}}

\bibliography{bib}

\allowdisplaybreaks

\title{Deep Partial Hedging}
\date{\today}
\author{Songyan Hou}
\address{ETH Zurich, Department of Mathematics}
\curraddr{}
\email{songyan.hou@math.ethz.ch}
\thanks{}

\author{Thomas Krabichler}
\address{Eastern Switzerland University of Applied Sciences, Centre for Banking \& Finance}
\curraddr{}
\email{thomas.krabichler@ost.ch}
\thanks{}

\author{Marcus Wunsch}
\address{Zurich University of Applied Sciences, Institute of Asset \& Wealth Management}
\curraddr{}
\email{marcus.wunsch@zhaw.ch}
\thanks{}

\begin{document}

\maketitle

\begin{abstract}
Using techniques from deep learning (cf.~\cite{buehler_deep_2019}), we show that neural networks can be trained successfully to replicate the modified payoff functions that were first derived in the context of partial hedging by~\cite{follmer_efficient_2000}. 
Not only does this approach better accommodate the realistic setting of hedging in discrete time, it also allows for the inclusion of transaction costs as well as general market dynamics. 
\end{abstract}

\section{Introduction}
In a complete market, the writer of an option can eliminate her risk entirely if she initiates a continuous hedging process with a capital position that equals the Black-Scholes price of the option. 
If the option writer decides to post strictly less capital to initiate the hedge, she will be exposed to shortfall risk. 
In this situation, she could try to maximize the probability of replicating the option payoff, a strategy named \textit{quantile hedging} by~\cite{follmer_quantile_1999}. 
Yet another strategy would be \textit{efficient hedging} (cf.~\cite{follmer_efficient_2000, follmer_stochastic_2016}), which has the advantage of taking into account the magnitude of the expected shortfall, which quantile hedging does not.

\subsection{Basic Model Setting}
$W_t$ shall denote a standard one-dimensional Brownian motion defined on the complete probability space $(\Omega, \mathcal{F}, \P)$, where $(\mathcal{F}_t)_{t \geq 0}$ is the augmentation of the natural filtration $\mathcal{F}_t^{W}=\sigma({W}_s; 0\leq s\leq t)$ for all $t\geq 0$. We consider a \textit{complete} market with a single geometric Brownian motion
\begin{align}\label{eq:xit}
    \d X_t &= X_t \left( \mu\d t + \sigma \d W_t \right), 
\end{align}
where the drift $\mu\in\R$ and the volatility $\sigma>0$ are constant.\\
Given the contingent claim's payoff function $H=H(X_T)$, we look for an admissible hedging strategy $(V_0, \, \pmb\xi)$ with
\begin{align*}
    V_t = V_0 + \int_0^t \xi_s \d X_s, 
\end{align*}
where $\xi$ is a predictable process with respect to the Brownian motion $W$ such that either 
\begin{itemize}
    \item $\P[V_T > H]$ becomes maximal in the context of quantile hedging, or 
    \item $\E[\ell ((H-V_T)_+)]$ becomes minimal in the context of efficient hedging. 
\end{itemize}
In this note, we consider loss functions of the form $\ell(x) = x^p/p$, $p\in \R_+$. 
The limiting case $p \downarrow 0$ is, in fact, identical to the quantile hedging problem. F{\"o}llmer and Leukert showed that quantile/efficient hedging is equivalent to delta-hedging options with certain modified payoffs, cf.~\cite[ ][Proposition~5.2, and Figure~1]{follmer_efficient_2000}. 

\section{Contribution of this Note}
In this research note, we show that deep neural networks can be trained to approximate closely the modified payoffs for efficient hedging with lower partial moments with $p>1$ derived theoretically by~\cite{follmer_efficient_2000, leukert_absicherungsstrategien_1999}. 
We stress that no other information is needed for this training but the underlying random environment, the capital amount corresponding to the initial hedge, and the  option's target payoff function. 
To the best of our knowledge, this is the first algorithmic approach to partial hedging whose optimization takes into account transaction costs. 
\begin{remark}
There is an important limitation to our approach that we encountered when $p$ lies in the unit interval $[0, 1]$. 
For risk preferences within this range, \cite{follmer_efficient_2000} found that the modified contingent claims to be replicated have a knock-out feature that makes their payoff profiles discontinuous. 
It appears that deep hedging struggles with detecting these discontinuous profiles. Overcoming the problem is apparently not straightforward.
On the other hand, when the modified payoff is continuous (e.g., if $p>1$), then deep partial hedging does work. 
\end{remark}

\section{Numerical Results} 
In this section, we present the numerical results by using deep partial hedging. 
We choose similar parameters as in~\cite{follmer_efficient_2000}; a risk-free rate $r = 0.7$, the maturity $T = 10$, Black-Scholes dynamics with $\mu = 0.08$, $\sigma = 0.3$, the initial stock price $S_0 = 100$, and a European call option with strike $\kappa = 110$. We choose $N = 100$ time steps for the discretization. 
This leads us to the discretized value process
\begin{equation}\label{eq_update1}
    V_{t} = V_{t-1} + K_{t}(S_{t} - S_{t-1}),\quad t = 1,\hdots, T,
\end{equation}
where $(K_{t})_{t=1}^{T}$ denotes the adapted hedging strategy. 

Let us set a bankruptcy bound $B < 0$. 
In order to incorporate $0$-admissibility, we modify the update in the following way: if updating with \eqref{eq_update1} results in $V_{t+1} < B$, we set
\begin{equation}
    V_{t+1}  = (V_{t} - B) \exp\Big(\pi_{t}(\mu \d t + \sigma \d W_{t})- \frac{\pi_{t}^{2}\sigma^{2} \d t}{2}\Big) + B,\qquad \pi_{t} = \frac{K_{t}S_{t}}{V_{t}}
\end{equation}
instead. 
If $V_{t} = B$, we claim bankruptcy and leave the market, i.e.,  $K_{h} = 0$ prevails for all $h\geq t$. With this modification, we make sure all strategies are $B$-admissible. In our numerical experiments, we set $B = -100$.
Then, we numerically minimize the loss consisting of three terms
\begin{equation}
    \mathcal{L}(K) = \mathcal{L}_{p}(K) + \mathcal{L}_{\text{cost}}(K) + \mathcal{L}_{\text{ad}}(K),
\end{equation}
where 
\begin{equation}
\begin{aligned}
    \mathcal{L}_{p}(K) &= \E\Big[\ell ((H-V_T)_+) \Big],\\
    \mathcal{L}_{\text{cost}}(K) &= c_{\text{cost}}\E\Big[\sum_{t=1}^{T-1}\lvert K_{t+1} - K_{t}\rvert \cdot S_{t}\Big],\\
    \mathcal{L}_{\text{ad}}(K) &= c_{\text{ad}}\E\Big[(-\min_{t}V_{t})_{+}\Big]
\end{aligned}
\end{equation}
for the hyperparameters $c_{\text{cost}}$ and $c_{\text{ad}}$. The minimization of the first term $\mathcal{L}_{p}$ is the primary objective of efficient hedging. The second term takes into account proportional transaction costs. The last term penalizes the deviation from the $0$-admissibility. 
We consider the so-called deep strategy 
\begin{equation}
    K_{t} = \mathcal{N}_{t}(S_{t})
\end{equation}
where $\mathcal{N}_{t}$ belongs to a certain class of neural networks $\mathcal{NN}$. Its structure comprises two hidden layers and each hidden layer consists of $21$ nodes. The detailed configuration can be found in the publicly available code\footnote{ \url{https://github.com/justinhou95/DeepHedging}}. We utilize Adam mini-batch training with a learning rate of $0.01$, cf.~\cite{DBLP:journals/corr/KingmaB14}. For different risk aversion levels $p \in \{1, 1.1, 2\}$, the following tables depict the performance line-up for deep hedging and discretized delta hedging respectively.

\begin{table}[h]
\begin{tabular}{|l|r|r|r|r|}
\hline
            & $\phantom{5}p = 1\phantom{.}$ & $p = 1.1$ & $\phantom{1}p = 2\phantom{.}$  \\ \hline
% deep hedge (train)  &   28.26    &    28.02   &  58.93   &    111.98       \\ \hline
deep hedge   &   \textbf{18.14}    &   \textbf{ 16.75}    &  \textbf{14.31}       \\ \hline
delta hedge &  19.64   &    36.69   &  32.63     \\ \hline
\end{tabular}
\vspace{1em}
\caption{Efficient hedging loss for different risk aversion levels without transaction costs}
\end{table}

\begin{table}[h]
\begin{tabular}{|l|r|r|r|r|}
\hline
            & $\phantom{5}p = 1\phantom{.}$ & $p = 1.1$ & $\phantom{1}p = 2\phantom{.}$  \\ \hline
% deep hedge with cost (train)  &  38.75 & 40.71 & 66.64 & 98.91       \\ \hline
deep hedge & \textbf{18.09} & \textbf{16.39} & \textbf{15.58}  \\ \hline
delta hedge&  26.45 & 44.01 & 37.84  \\ \hline
\end{tabular}
\vspace{1em}
\caption{Efficient hedging loss for different risk aversion levels with proportional transaction costs of $1\%$}
\end{table}

Finally, we compare the terminal wealth between deep hedge and delta hedge, from which we can see that the violation of admissibility by the delta hedge is substantial. 

\begin{figure}[H]
\includegraphics[width=\linewidth]{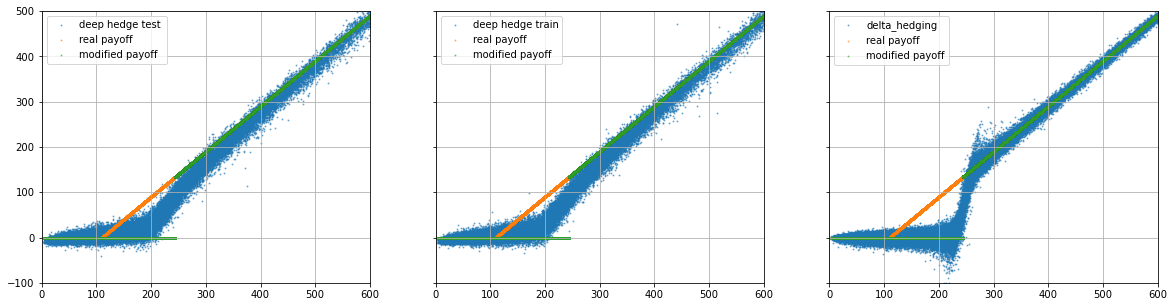}
\caption{Terminal wealth ($p=1$)}
\end{figure}

\begin{figure}[H]
\includegraphics[width=\linewidth]{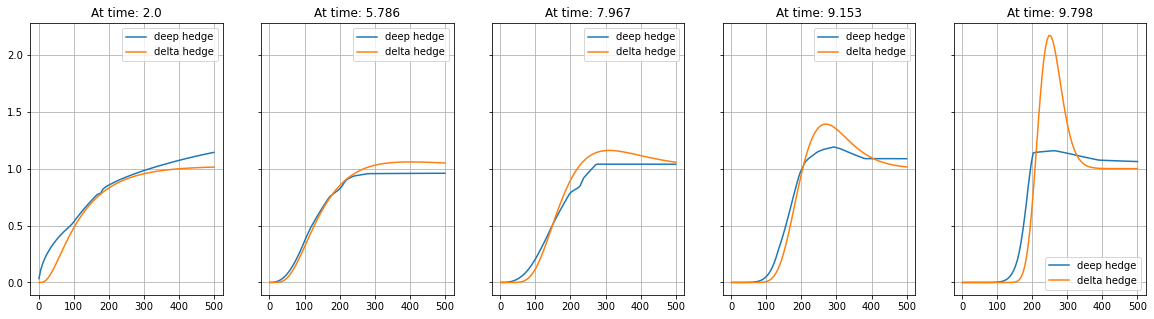}
\caption{Hedging strategy ($p=1$)}
\end{figure}

\begin{figure}[H]
\includegraphics[width=\linewidth]{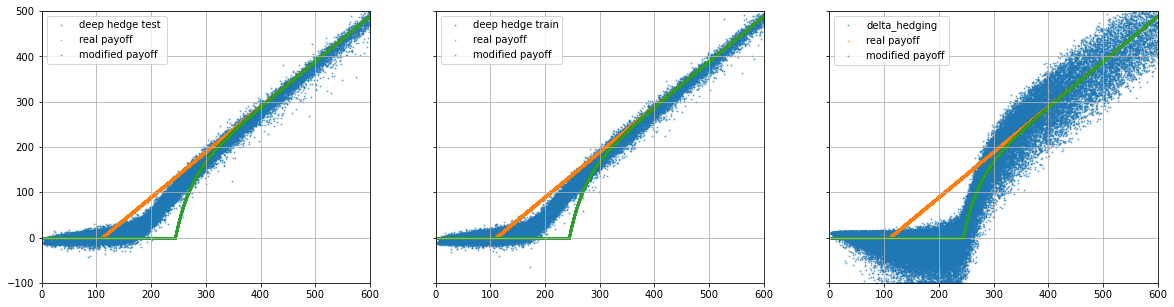}
\caption{Terminal wealth ($p=1.1$)}
\end{figure}

\begin{figure}[H]
\includegraphics[width=\linewidth]{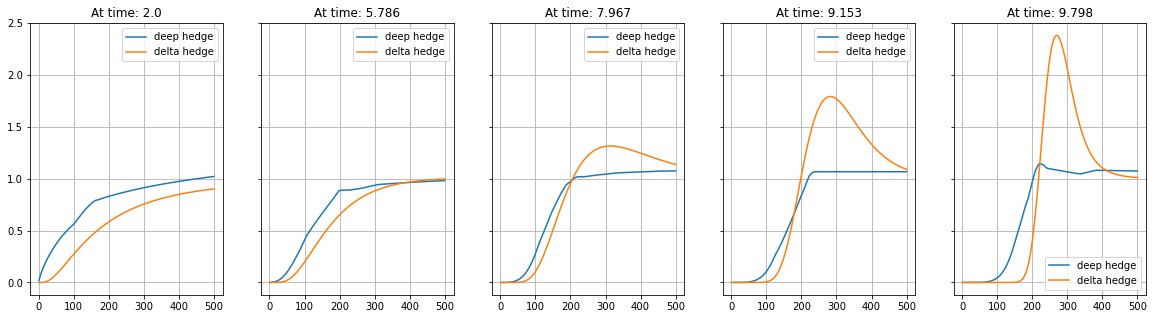}
\caption{Hedging strategy ($p=1.1$)}
\end{figure}

\begin{figure}[H]
\includegraphics[width=\linewidth]{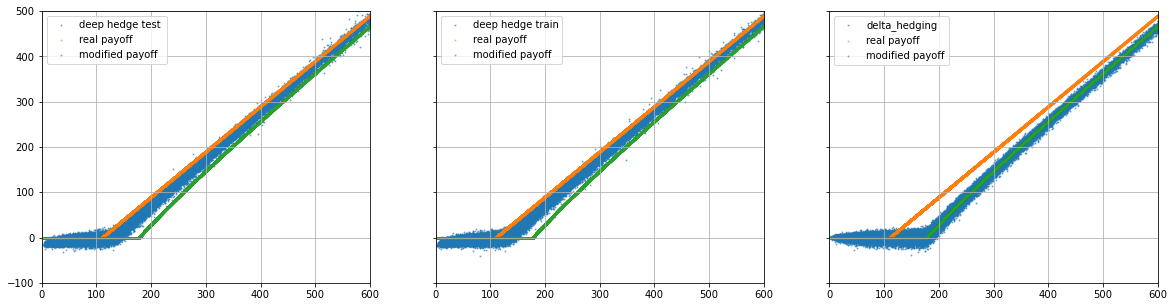}
\caption{Terminal wealth ($p=2$)}
\end{figure}

\begin{figure}[H]
\includegraphics[width=\linewidth]{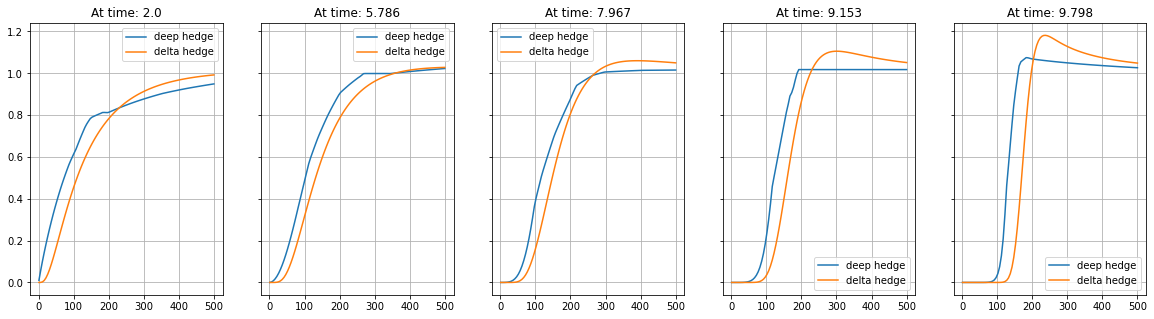}
\caption{Hedging strategy ($p=2$)}
\end{figure}

\section{Conclusion \& Outlook} 
In this research note, we have shown that deep partial hedging can replicate the modified contingent claims as derived theoretically in the context of efficient hedging by~\cite{follmer_efficient_2000}, without any prior knowledge of the latter's payoff profile. 
Our findings hold for risk-neutral and risk-averse option writers ($p\ge1$); for risk-taking option writers ($p\in [0, 1)$), it appears to be difficult to capture the modified payoff with two barriers using deep neural networks (cf.~\cite{follmer_efficient_2000}). 
We defer further research into these challenges, as well as the investigation of deep partial hedging for more general market dynamics and derivatives, to future work. 

\printbibliography[title={References}]

\end{document}